\documentclass[12pt,a4paper]{article}
\pdfoutput=1
\usepackage[utf8]{inputenc}
\usepackage[]{graphicx}
\usepackage{amsmath}
\usepackage{amssymb}
\usepackage[]{color}
\usepackage{subcaption}
\usepackage{natbib}
\usepackage{xspace}
\usepackage{calc}
\usepackage[noae]{Sweave}
\usepackage{authblk}
\usepackage{algorithm}
\usepackage{algpseudocode}
\newcommand{\ie}{\emph{i.e.}\@\xspace}

\title{Sequential rank agreement methods for comparison of ranked lists}
\author{Claus Thorn Ekstrøm}
\author{Thomas Alexander Gerds}
\author{Andreas Kryger Jensen}
\author{Kasper Brink-Jensen}
\affil{Biostatistics, University of Copenhagen}

\begin{document}


\maketitle

\begin{abstract}
  The comparison of alternative rankings of a set of items is a
  general and prominent task in applied statistics. Predictor
  variables are ranked according to magnitude of association with an
  outcome, prediction models rank subjects according to the
  personalized risk of an event, and genetic studies rank genes
  according to their difference in gene expression levels. This
  article constructs measures of the agreement of two or more ordered
  lists. We use the standard deviation of the ranks to define a
  measure of agreement that both provides an intuitive interpretation
  and can be applied to any number of lists even if some or all are
  incomplete or censored. The approach can identify change-points in
  the agreement of the lists and the sequential changes of agreement
  as a function of the depth of the lists can be compared graphically
  to a permutation based reference set.  The usefulness of these tools
  are illustrated using gene rankings, and using data from two Danish
  ovarian cancer studies where we assess the within and between
  agreement of different statistical classification methods.
\end{abstract}


\section{Introduction}

Ranking of items or results is common in scientific research and
ranked lists occur naturally as the result of many statistical
applications. Regression methods rank predictor variables according to
magnitude of their association with an outcome, prediction models rank
subjects according to their risk of an event, and genetic studies rank
genes according to their difference in gene expression levels across
samples.

When several rankings of the same items are available, a common
research question is to what extent they agree. In particular, is it
possible to identify an optimal rank until which the lists agree on
the items?  A typical situation arises in high-dimensional genomics
studies when several analysis methods are applied to rank a list of
genes according to their association with a phenotype, treatment
effect or other outcome. A measure of agreement of gene rankings
obtained by different methods could often help to identify which genes
are worth to pursue in further experiments.

Several approaches exist to measure the ``distance'' between two
ranked lists. Of these, Kendall's $\tau$ \citep{kendall1948} and
Spearman’s footrule $\rho$ \citep{Spearman1910} are among the most
well-known, but these measures do not distinguish between agreement in
the top versus towards the bottom of the lists and the provide a
single measure of the overall distance. \citet{Shieh1998} proposed a
weighted version of Kendall's $\tau$ where each pair of rankings can
be assigned different weights, and \citet{Yilmaz2008} proposed the
$\tau_{ap}$ which places higher emphasis on the top of the lists.
Spearman’s footrule uses the ranks of the variables for calculation of
the distance and the use of ranks are also employed in the $M$ measure
of \citet{Bar-Ilan2006} where the reciprocal rank differences are used
to calculate the similarity measure.

Other recent approaches consider the intersection of lists as the basis
for a similarity measure. However, simple intersection also places
equal weights on all depths of the list and therefore \citet{Fagin2003} and
\citet{Webber2010} proposed weighted intersections which put more
emphasis on the top of the lists. One proposal is denoted the average
overlap (AO) and used for comparison later on. Specifically,
\citet{Webber2010} define their rank-biased overlap (RBO) by weighting
with a converging series to ensure that the top is weighted higher
than the potentially non-informative bottom of the lists.
It is possible to use the existing methods to calculate agreement of
lists until a given depth, i.e., limited to the $d$ items of each
list. However, the interpretation may not be straightforward,
especially in the case of more than two lists, and they may not
accommodate partial rankings.

In this article we introduce sequential rank agreement for measuring
agreement among (partially) ranked lists.  The general idea is to
define agreement based on the sequence of ranks from the first $d$
elements in each list. As agreement metric we adapt the limits of
agreement known from agreement between quantitative variables
\citep{alt:bland:1983,Carstensen2010}. Our proposed approach allows us
to compare more than two lists simultaneously, it provides a dynamic
measure of agreement as a function of the depth in the lists, places
high weight on the top of the list, accommodates censored/incomplete
lists of varying lengths, and has a natural interpretation that
directly relates to the ranks. Graphical illustration of sequential
rank agreement potentially allows us to infer a change-point, i.e., a
list depth where a change in the agreement of the lists occur but we
also provide randomization-based graphical tools to compare the
observed rank agreement to the expected agreement found in
non-informative data.  In this sense our approach is a combination and
generalization of some of the ideas of \citet{Carterette2009} and
\citet{Boulesteix2009}. The former compares two rankings based on the
distance between them as measured by a multivariate Gaussian
distribution and the latter presents an overview of approaches for
aggregation of ranked lists including bootstrap and leave-one-out
jackknife approaches.

The manuscript is organized as follows: In the next section we define
sequential rank agreement for multiple ranked lists and discuss how to
handle incomplete/censored/partial lists. In section 3 we present and
discuss approaches to evaluate the results obtained from sequential
rank agreement. Finally we apply the sequential rank agreement to two
Danish ovarian cancer studies before we discuss the findings along
with possible extensions. The approaches presented in this manuscript
are available in the R package \texttt{SuperRanker}.

\section{Methods}

Consider a set of $P$ different items $X=\{X_1,\dots,X_P\}$ and a
ranking function $R: \{X_1,\dots,X_P\}\to \{1,\dots,P\}$, such that
$R(X_p)$ is the rank of item $X_p$. The inverse mapping $R^{-1}$ gives
the item $R^{-1}(r)$ that was assigned to rank $r\in\{1,\dots,P\}$. An
ordered list is the realization of a ranking function $R$ applied to a
data sample $X$ and we let $R_1(X),\dots,R_L(X)$, $L\geq2$, denote
ordered lists obtained from $L$ different ranking functions applied to
the same data $X$. We will use the same notation for the equivalent
scenario where we have ordered lists obtained from applying the same
ranking function to $L$ samples from the same population, i.e., for
ranked lists $R(X^1), \ldots, R(X^L)$.  Panels (a) and (b) of
Table~\ref{tab:example} show a schematic example of these
mappings. Thus if $R_l^{-1}(1)=X_{34}$ then item $X_{34}$ is ranked
first in list $l$ and similarly $R_l(X_{34})=1$.

\begin{table}[tb]
  \caption{Example set of ranked lists. (a) shows the ranked list of
    items for each of three lists, (b) presents the ranks obtained by
    each item in each of the three lists and (c) shows the cumulative
    set of items up to a given depth in the three lists.}
\begin{center}
\footnotesize{
  \begin{subtable}{4.5cm}%
    \caption{}
      \begin{tabular}{cccc}
        \hline\hline 
        Rank & $R^{-1}_1$ & $R^{-1}_2$ & $R^{-1}_3$ \\ \hline
        1 & A & A & B \\
        2 & B & C & A \\
        3 & C & D & E \\
        4 & D & B & C \\
        5 & E & E & D \\ \hline
    \end{tabular}
  \end{subtable}
\hspace{1em}
  \begin{subtable}{4cm}%
    \caption{}
    \begin{tabular}{cccc} \hline\hline
    Item & $R_1$ & $R_2$ & $R_3$ \\ \hline
    A & 1 & 1 & 2 \\
    B & 2 & 4 & 1 \\
    C & 3 & 2 & 4 \\
    D & 4 & 3 & 5 \\
    E & 5 & 5 & 3 \\ \hline
  \end{tabular}
\end{subtable}
\hspace{1em}
\begin{subtable}{4cm}%
  \caption{}
\begin{tabular}{cc} \hline\hline
Depth  &  $S_d$ \\ \hline
1 & $\{$A, B$\}$\\
2 & $\{$A, B, C$\}$ \\
3 & $\{$A, B, C, D, E$\}$ \\
4 & $\{$A, B, C, D, E$\}$ \\
5 & $\{$A, B, C, D, E$\}$ \\ \hline
\end{tabular}
\end{subtable}
}
\end{center}
\label{tab:example}
\end{table}

The agreement of the lists regarding the rank given to an item $X_p$ can be
measured by
\begin{equation}
  A(X_p) = f(R_1(X_p), \ldots, R_L(X_p)),
\end{equation}
for a non-negative real-valued distance function $f.$ Throughout this
paper we will use the sample standard error as our function $f$ and
hence use
$$A(X_p) = \sqrt{\frac{\sum_{i=1}^L (R_i(X_p) - \bar{R}(X_p))^2}{L-1}},
$$
where $\bar{R}(X_p)$ is the average rank assigned to item $X_p$, but
other possibilities will be discussed below. For each item, the sample
standard error has an interpretation as the average distance of the
individual rankings from the average ranking over all the lists, and
it corresponds to the same measure that is used in method comparison
studies to compute the limits of agreement \citep{alt:bland:1983}.

For an integer $1\le d\le P$ we define the unique set of items found
by merging the first $d$ elements of each of the $L$ lists, i.e., the
set of items ranked less than or equal to $d$ in any of the lists:
\begin{equation}
S_d = \{R_l^{-1}(r) ; r \leq d, l = 1, \ldots, L \},
\end{equation}
which is exemplified in Panel (c) of Table
\ref{tab:example}.

The \emph{sequential rank agreement} is the pooled standard
deviation of the items found in the set $S_d$:
\begin{equation}
\textrm{sra}(d)= \sqrt{\frac{\sum_{\{p \in
      S_d\}}(L-1)A(X_p)^2}{(L-1)|S_d|}}, \label{def:sra}
\end{equation}
where $|S_d|$ refers to the cardinality of the set $S_d$. Values of
\textrm{sra} close to zero suggests that the lists agree while larger
values suggest disagreement. Just like in method comparison studies we
prefer smaller limits of agreement because that means that the
differences between the methods is small. If the ranked lists are
identical then the value of sequential rank agreement will be zero for
all depths $d$. The sequential rank agreement can be interpreted as
the average distance of the individual rankings of the lists from the
average ranking for each of the items we have seen until depth $d$.

Note that the terms that are part of \eqref{def:sra} are not
independent since every list contains the ranks from 1 to $P$ exactly
once. As a consequence we simply use the pooled standard deviation as
a measure of the average distance between rankings and we refrain from
using traditional distributional results for this measure.

\subsection{Agreement among fully observed lists}
\label{sec:amfol}

The simplest case occurs when all $L$ lists are fully observed, \ie,
we have observed the rank of all $P$ items for all $L$ lists. Fully
observed ranked lists are common and arise, for example, when
different statistical analysis methods are applied to a single dataset
to produce lists of predictors ranked according to their importance,
or if the same analysis method is applied to data from different
populations.

\begin{figure}[tb]
   \begin{center}
 \includegraphics[width=.49\textwidth]{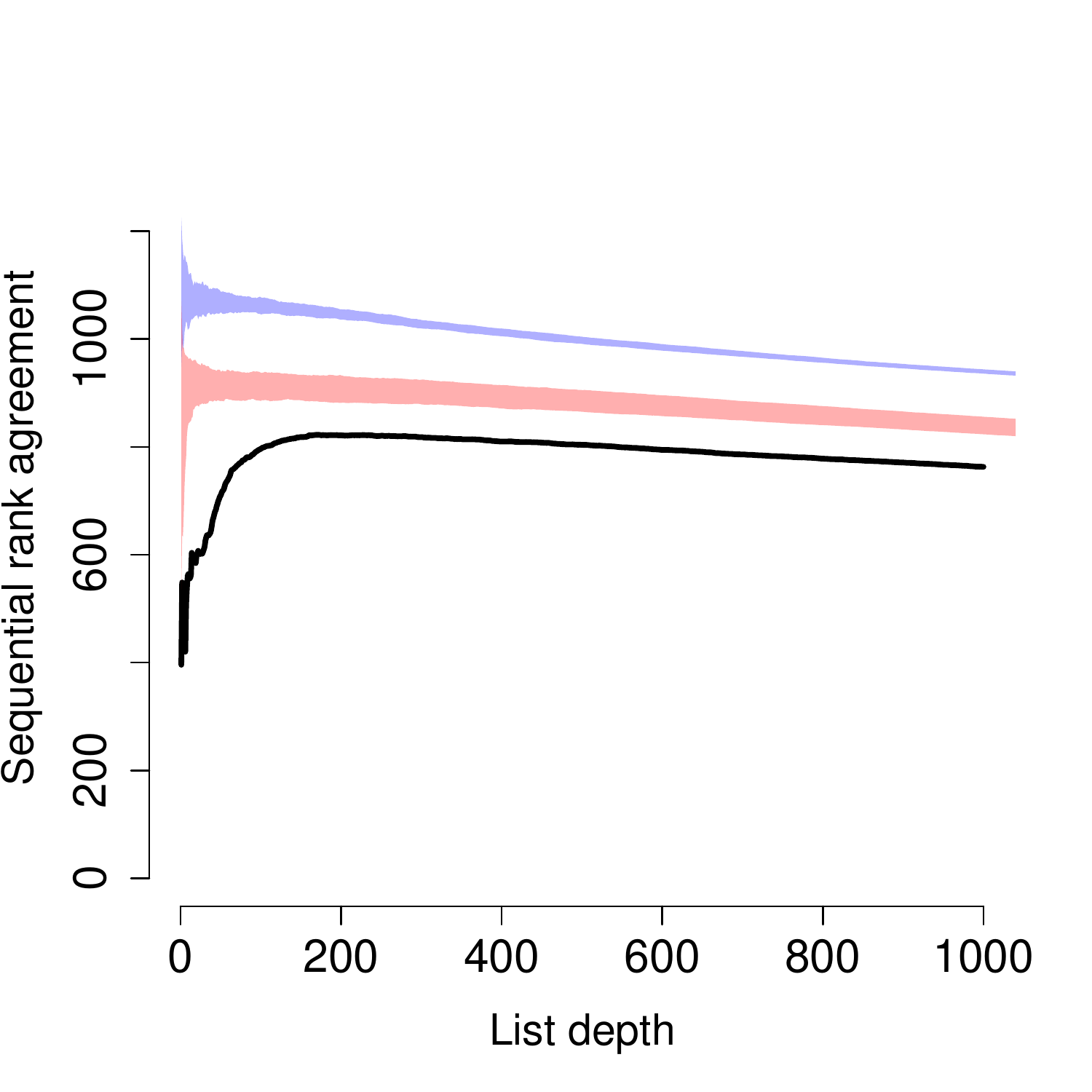}
 \includegraphics[width=.49\textwidth]{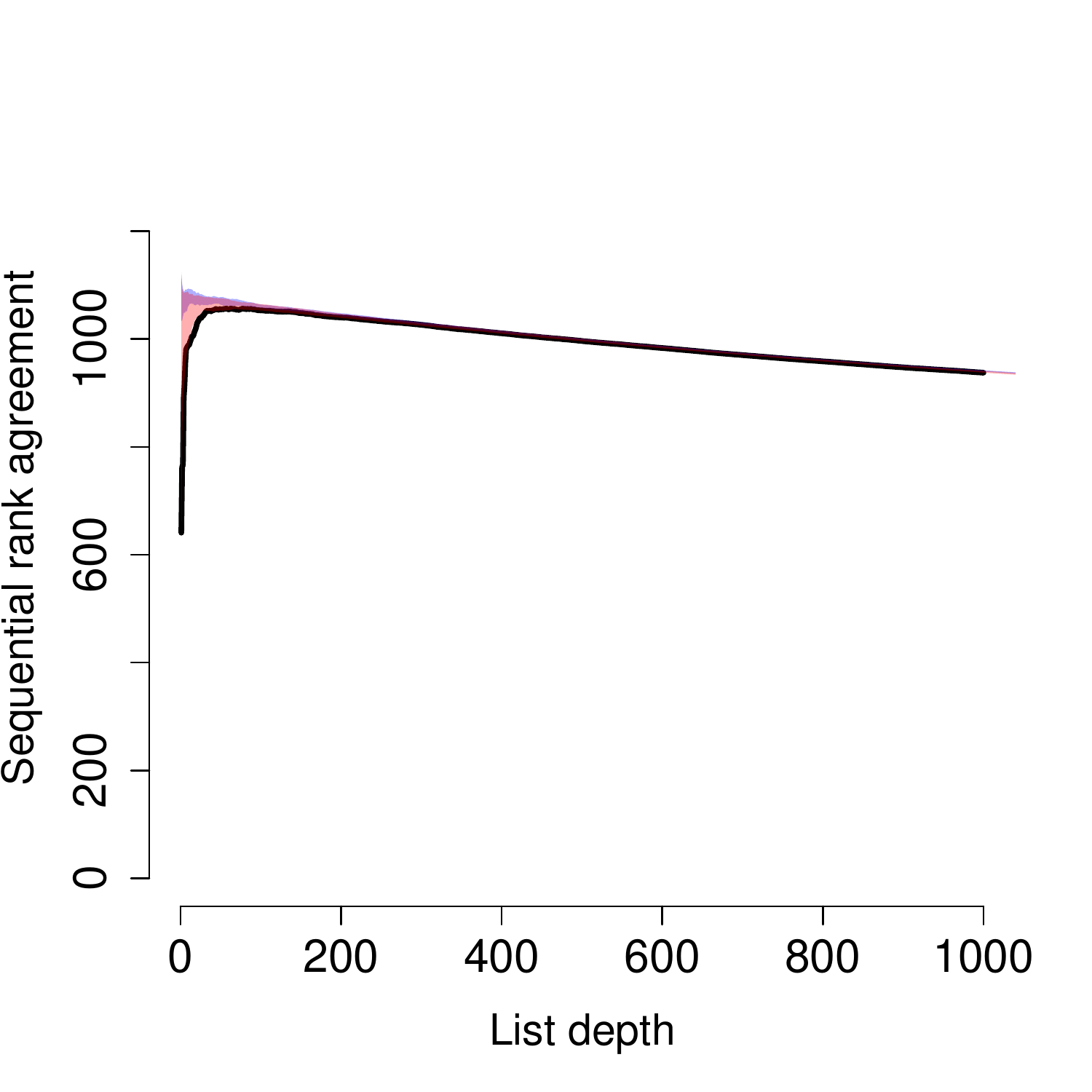}
 \includegraphics[width=.49\textwidth]{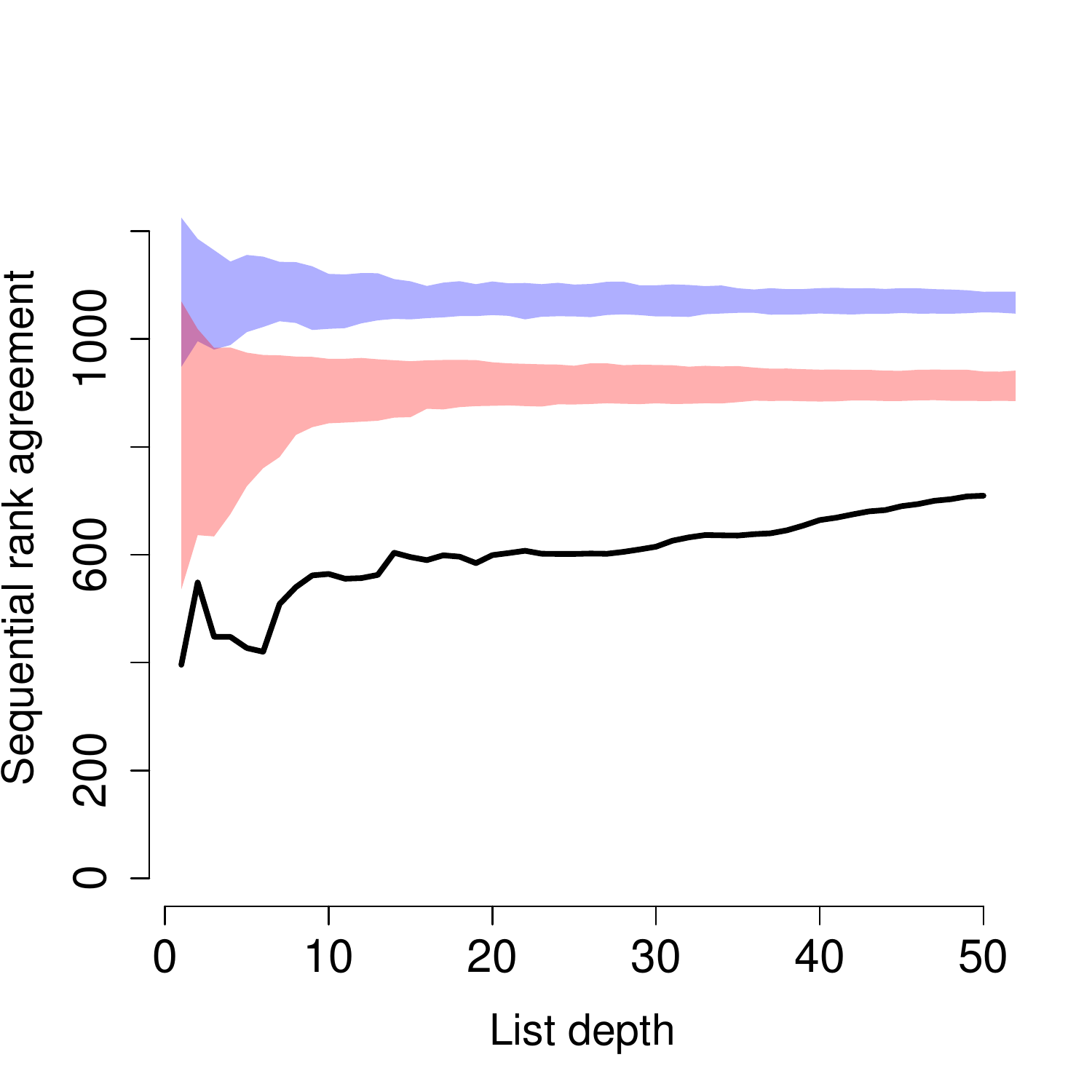}
 \includegraphics[width=.49\textwidth]{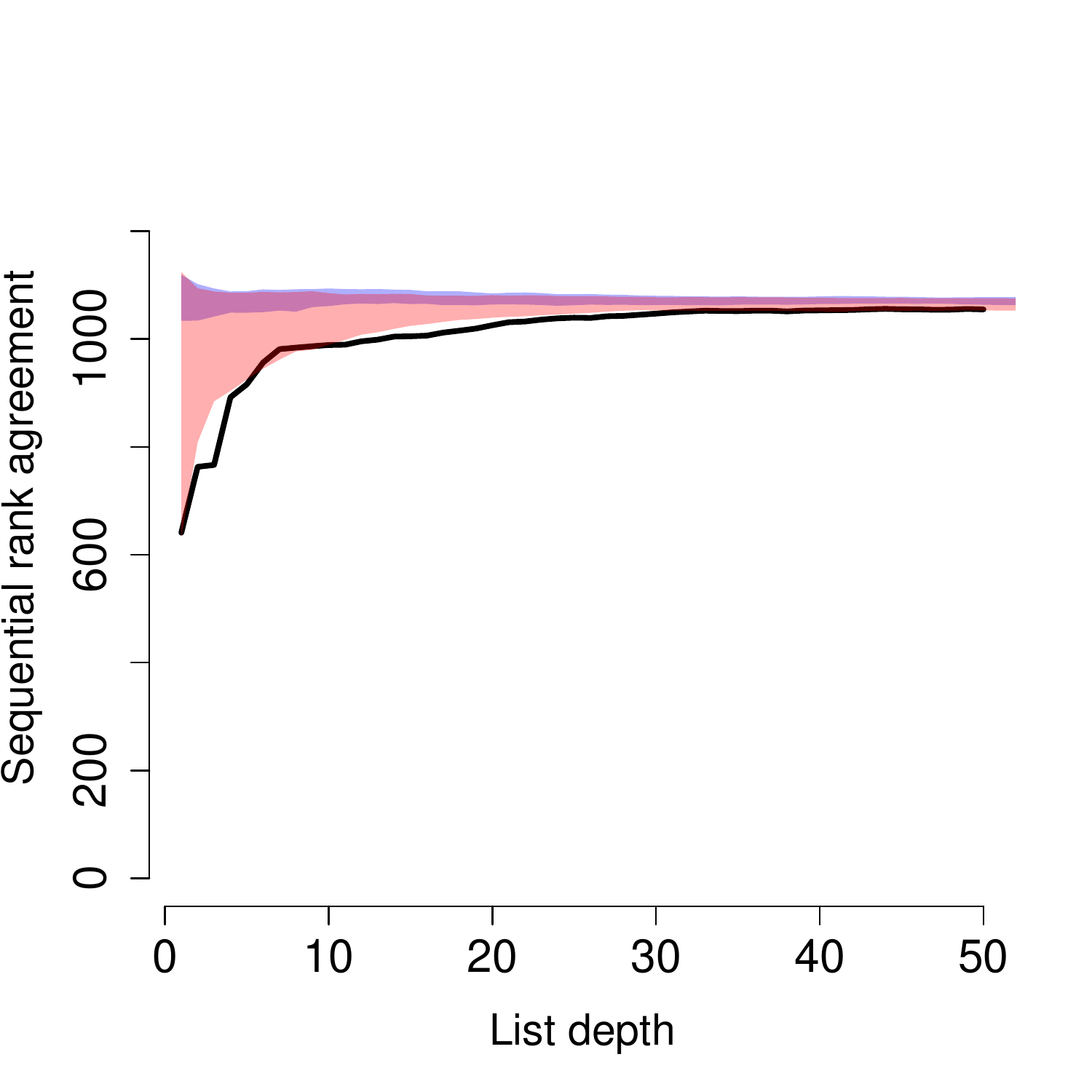}
 \end{center}
 \caption{Left panels: Sequential rank agreement for 4 different
   analysis methods applied to the 3051 genes in the Golub data /black
   line). Right
   panels: Corresponding sequential rank agreement for the same data
   but where only the top 20 ranked items are available. The blue and
   red areas correspond to the independent and randomized reference
   hypothesis areas, respectively. The bottom plots are identical to
   the corresponding top plots but have been zoomed in on the first
   part of the $x$ axis.}
 \label{fig:example1}
 \end{figure}

With fully observed lists we can plot the sequential rank agreement
\eqref{def:sra} as a function of depth $d$. An example is seen in top
panels of Figure~\ref{fig:example1} where four analysis methods were
used to rank 3051 gene expression values measured on 38 tumor mRNA
samples in order to classify between acute lymphoblastic leukemia
(ALL) and acute myeloid leukemia (AML)
\citep{Golub1999}. Preprocessing of the gene expression data was done
as described in \citet{Dudoit2002} and the four different analysis
approaches were: marginal two-sample $t$ tests, marginal logistic
regression analyses, elastic net logistic regression
\citep{friedman2010regularization}, and marginal maximum information
content correlations (MIC) \citep{Reshef2011}. For the first two
methods, the genes were ranked according to minimum $p$ value, for
logistic regression the genes were ordered by size of the
corresponding coefficients (after standardization), and MIC was
ordered by absolute correlation which resulted in the top rankings
seen in Table~\ref{tab1}.

The sequential rank agreement plotted in Figure~\ref{fig:example1}
shows how many ranks apart we on average expect the genes found in the
top $d$ part of the lists to be for each depth $d$.  It can be seen
that the sequential rank agreement limits are better towards the top
of the lists (smaller values on the $y$ axis) corresponding to
\emph{better} agreement than in the tails of the lists. It is clear
from the curve in Figure~\ref{fig:example1} that there is a
substantial drop in agreement even after the first depth. Thus, if we
were to restrict attention to a small set of predictors then we would
focus on predictors 2124, 829, and 378 as seen in Table~\ref{tab1}.

\begin{table}[tb]
\centering
\caption{Top 10 list of ranked results from the Golub data. Numbers indicate the predictor/gene for the given ranking and method} 
\label{tab1}
\begin{tabular}{rrrrr}
  \hline
Ranking & T & LogReg & ElasticNet & MIC \\ 
  \hline
1 & 2124 & 2124 & 829 & 378 \\ 
  2 & 896 & 896 & 2124 & 829 \\ 
  3 & 2600 & 829 & 2198 & 896 \\ 
  4 & 766 & 394 & 1665 & 1037 \\ 
  5 & 829 & 766 & 1920 & 2124 \\ 
  6 & 2851 & 2670 & 1042 & 808 \\ 
  7 & 703 & 2939 & 808 & 108 \\ 
  8 & 2386 & 2386 & 849 & 515 \\ 
  9 & 2645 & 1834 & 937 & 2670 \\ 
  10 & 2002 & 378 & 1995 & 2600 \\ 
   \hline
\end{tabular}
\end{table}

 If the lists agree on all items then the sequential rank agreement is
 zero for all depths. Generally, if there are changes in the level of
 rank agreement then this suggests that there are sets of items that
 are ranked similarly in all lists and other sets of items that have
 been assigned vastly different ranks in the lists. In gene
 association studies we generally expect the agreement among the lists
 to be better towards the top of the lists and worse towards the end
 of the lists. In this case the sequential rank agreement curve will
 start at a low level and then increase until it levels off as in
 Figure~\ref{fig:example1}.

\subsection{Analysis of partial/censored lists}

Incomplete or partial lists are also a common occurrence. They arise
for example in case of missing data (items), when comparing top $d$
list results from publications, or when some methods only rank a
subset of the items. For example, penalized regression based on the
Lasso provides a sparse set of predictors that have non-zero
coefficients. There is no obvious ordering of the set of predictors
whose coefficient has been shrunk to zero and thus we end up with a
partial ordering. Incomplete lists also occur when the analyst decides
to censor all lists at depth $d$ such that the measurement of
agreement is restricted to the first $d$ elements of each list, or
when the analyst only wants to consider the ranking of items that have
been found to be statistically significant.

Sequential rank agreement can be generalized to partial lists in the
following way. Let $\Lambda_l\subset X$ be the subset of $d_l$ items
that have been ranked highest in list $l$.  The case where all lists are
censored at the same depth $d$ corresponds to $d_1=\cdots=d_L=d$. For
incomplete/censored lists the rank function becomes
\begin{equation}
  \widetilde R_l(X_p) = \left\{\begin{array}{cl} \{R_l(X_p)\} & \text{ for } X_p\in \Lambda_l \\
      \{d_l+1,\dots,P\} & \text{ for } X_p \not\in \Lambda_l\end{array}\right.
\end{equation}
where we only know that the rank for the unobserved items in list $l$
must be larger than the largest rank observed in that list.

The agreement, $A(X_p)$, cannot be computed directly for all
predictors in the presence of censored lists because the exact rank
for some items will be unknown. Also, recall that the rankings within
a single list are not independent since each rank must appear exactly
once in each list. Thus, we cannot simply assign the same number
(e.g., the mean of the unassigned ranks) to the censored items since
that would result in less variation of the ranks and hence less
variation of the agreement, and it would artificially introduce a
(downward) bias of agreement for items that are censored in multiple
lists.

Instead we randomize the ranks $\{d_{l}+1,\dots,P\}$ to the items that
do not occur in list $\Lambda_l$. One realization of the $L$ rankings
of the set $X$ is obtained by randomizing the missing items of each
list. By randomizing a large number of times we can compute
\eqref{def:sra} for each realization, and then compute the sequential
rank agreement as the pointwise (for each depth) average of the rank
agreements. The algorithm is described in detail in
Algorithm~\ref{sra-algorithm}.

\begin{algorithm}
\caption{Sequential rank agreement algorithm for censored lists}
\label{sra-algorithm}
\begin{algorithmic}[1]
  \Procedure{Censored rank agreement}{} \State Let $B$ be the number
  of permutations to use \For{each $b \in B$} \For{each censored list
    $l \in L$}
  \State \parbox[t]{\dimexpr\linewidth-\algorithmicindent-1.7cm}{Permute
    the unassigned ranks, $\{d_l+1, \ldots, P\}$, and assign them
    randomly to the items \emph{not} found in the list, i.e.,
    $\Lambda^\complement=X\setminus\Lambda_l$. Combine the
    result with $\Lambda_l$ to fill out the list.}
\EndFor
\State \parbox[t]{\dimexpr\linewidth-\algorithmicindent-1cm}{Let sra($b$) be the sequential rank agreement computed from the
filled out lists.}
\EndFor
\State Return element-wise averages across all $B$ permutations of sra($b$).
\EndProcedure
\end{algorithmic}
\end{algorithm}

The proposed approach is based on two assumptions: 1) that the most
interesting items are found in the top of the lists, and 2) that the
censored rankings provide so little information that it is reasonable
to assume that they can be represented by a random order. The first
assumption is justifiable because we have already accepted that it is
reasonable to rank the items in the first place. The second assumption
is fair in the light of the first assumption provided that we have a
``sufficiently large'' part of the top of the lists available.

When the two assumptions are satisfied then it is clear that the
interesting part of the sequential rank agreement curves is restricted
to the depths where the number of censored items is
low. Generally, without additional prior knowledge about the
distribution of the censored items it seems reasonable to restrict the
attention of the sequential rank agreement to depths smaller than
$\max(d_1, \ldots, d_L)$ where at least one list provides an
uncensored item.

Like for fully observed lists we generally expect the sequential rank
agreement to start low and then increase unless the lists are
completely unrelated (in which case the sequential rank agreement will
be constant at a high level) or if the lists mostly agree on the
ranking (in which case the sequential rank agreement will also be
constant but at a low level). For partially ranked lists we also
expect a change-point around the depth where the lists are
censored. This is an artefact stemming from the fact that we assume
that the remainder of the lists can be replaced by a simple
permutation of the missing items.

\section{Evaluating sequential rank agreement}

To interpret the sequential rank agreement values we propose two
different benchmark values corresponding to two different
hypotheses. We wish to determine if we observe better agreement than
what would be expected if there were no relevant information available
in the data.

The first reference hypothesis is
\begin{eqnarray*}
H_0  & : &  \text{The list rankings correspond to complete randomly}\\
       &  & \text{permuted lists},
\end{eqnarray*}
which not only assumes that there is no information in the data on
which the rankings are based but also that the methods are completely
independent.

Alternatively, we can remove the restriction on the independence among
the methods by only requiring that there is no information in the
ranking but that the rankings are all based on applying the
method/approaches to the same data
\begin{eqnarray*}
\widetilde H_0 & :&  \text{The list rankings are based on data containing}\\
& &   \text{no association to the outcome.}
\end{eqnarray*}
$H_0$ is a quite unrealistic null hypothesis but we can easily obtain
realizations from that null hypothesis simply by permuting the items
within each list and then computing the sequential rank agreement for
the permuted lists. In the fully observed case each experiment
contains $L$ lists of random permutations of the items in $X$. For the
censored/partially observed case we first permute the items
$X_1,\dots,X_P$ and then censor list $l$ at $d_l$. The sequential rank
agreement curve from the the original lists can then be compared to,
say, the pointwise 95\% quantiles of the observed rank agreements
obtained under $H_0$.

To obtain the distribution under $\widetilde H_0$ the idea is to
repeat the ranking procedures for unassociated data many times.  Thus,
we first permute the outcome variable of the data set on which the
rankings are based. This removes any association between the predictor
variables and the outcome. Then we apply the methods to the permuted
data set to generate $L$ new rankings and subsequently compute the
sequential rank agreement. Note that we only permute the outcomes and
preserve the structure of the candidate predictors. The randomization
approach requires that we have the original data available and as such
it may not be possible to evaluate $\widetilde H_0$ in all situations.

If the sequential rank agreement for the original data lies
substantially below the distribution of the sequential rank agreements
obtained under either $H_0$ or $\widetilde H_0$ then this suggests that
the original ranked lists agree \emph{more} than what we would expect
in data with no information, and therefore that the information in the
lists is significantly more in agreement than what would be expected.

Figure~\ref{fig:example1} shows the empirical distributions of
sequential rank agreement under $H_0$ and $\widetilde H_0$ each based
on $400$ permutations of the Golub data described in Section
\ref{sec:amfol}. Not surprisingly, the sequential rank agreement under
$\widetilde H_0$ is lower than the sequential rank agreement under
$H_0$ because the four methods used to rank the data ($t$ test,
logistic regression, elastic net, and MIC) generally tend to identify
(and rank) similar predictors even if there are only spurious
associations. The two bottom panels in Figure~\ref{fig:example1} also
indicate that the observed sequential rank agreement (the black line)
is better than what would be expected by chance for data that contain
no information since it lies below the reference areas. However, the
censored data also suggests that there may be at most 1 or 2 ranked
items towards the top of the lists that yield a result better than
what would be expected (the bottom-right plot).

It is important to stress that neither $H_0$ nor $\widetilde H_0$ are
related to questions regarding the association between the outcome and
the predictors in the data set. Both hypotheses are purely considering
how the rankings agree in a situation where there is no relevant
information available in the data used for creating the rankings.

\section{Comparison to average overlap}

The average overlap is a measure that is widely used in comparisons of
two fully observed ranked lists \citep{Fagin2003,Webber2010} and the
idea behind the average overlap closely resembles the sequential rank
agreement. In this section we present an extension of the average
overlap to more that two lists and compare this to sequential rank
agreement.

The overlap among $L$ lists observed until depth $d$ is defined as the
proportion of items that are found in all $L$ lists compared to the
possible number of items found in all lists observed until depth $d$:
\begin{equation}
  O(d) = \frac{| \cap_{l=1}^L \{R_l^{-1}(r); r\leq d \} |}{d}. \label{def:overlap}
\end{equation}
Typically the overlap is only considered for pairwise comparisons
\citep{Bar-Ilan2006,Boulesteix2009} but formula \eqref{def:overlap}
is easily extended to the situation where $L>2$ as shown above.

The average overlap at depth $d$ is defined as the average of the
overlaps until depth $d$,
\begin{equation}
AO(d) = \frac1d\sum_{i=1}^d O(d).
\end{equation}
This definition directly ensures that further emphasis is put on items
in the top of the lists since their item specific overlap contributes
to the calculation of average overlap at higher depths.

There are three main drawbacks of the average overlap method: First it
is highly sensitive to single items in the top of one the lists.  The
average overlap profile may change dramatically depending on the
presence or absence of similar items in the beginning of the
lists. For example, if all lists have the same item at the top then
the average overlap attains its maximum at 1, but if just one of the
lists differ then the overage overlap starts at its minimum of
0. Secondly, when the number of lists, $L$, is high then it might be
difficult to obtain a non-zero overlap (and hence a non-zero average
overlap) at the beginning of the lists because the overlap will be
zero until an item is present in all $L$ lists. Finally, the average
overlap has an interpretation in terms of moving averages of
percentages (of $L$ sets) which is somewhat less intuitive than our
proposed sequential rank distance.

In the following we compare the average overlap and the sequential
rank agreement using the analysis of the Golub data
\citep{Golub1999}. We consider a typical application of the average
overlap and thus restrict attention to only two gene rankings obtained
by the $t$ tests and logistic regression. These two marginal modeling
approaches should result in very similar gene rankings. The first two
columns of Table~\ref{tab1} show the actual top-10 ranking of the
predictors.

\begin{figure}[tb]
\begin{center}
\includegraphics[width=.49\textwidth]{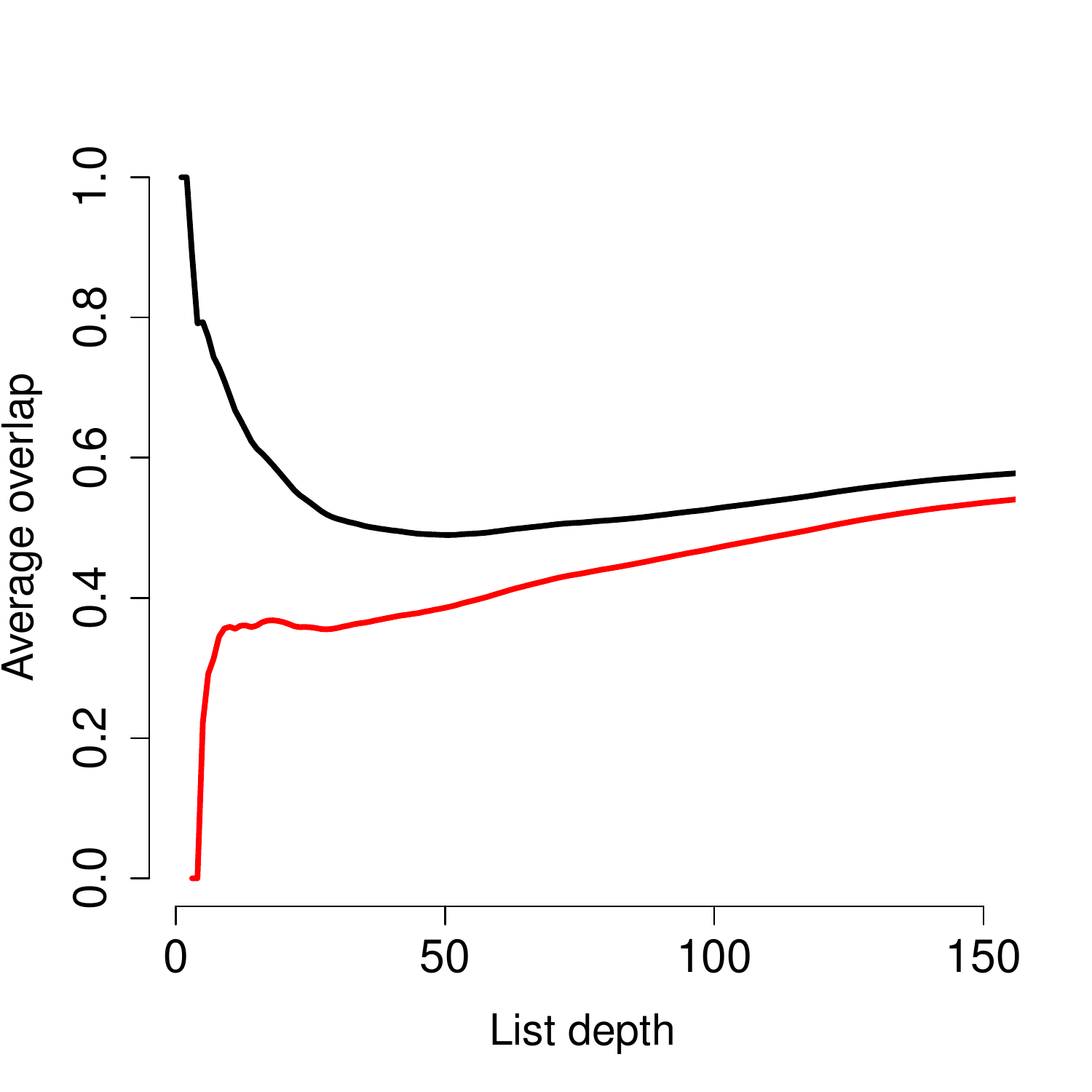}
\includegraphics[width=.49\textwidth]{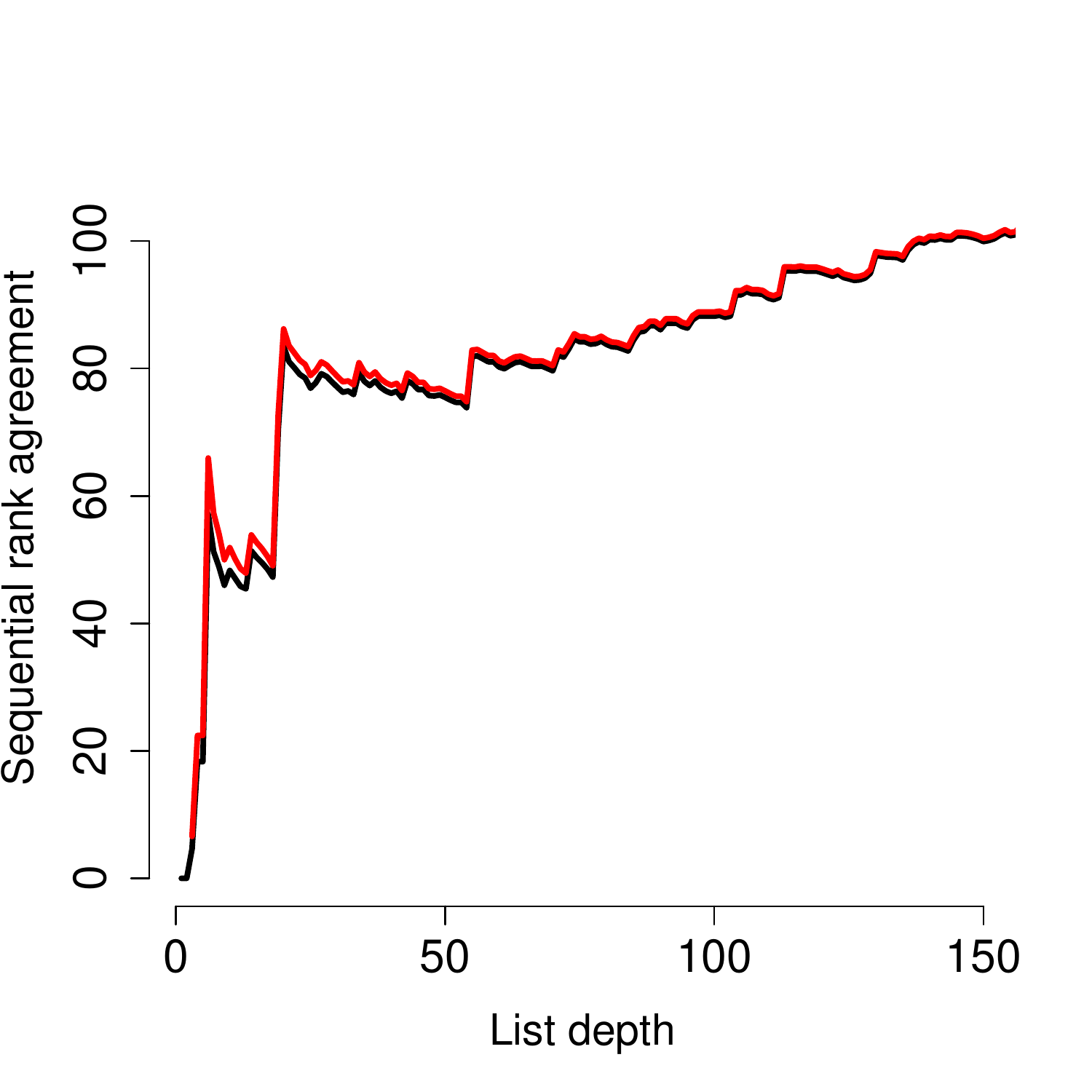}
\end{center}
 \caption{Average overlap (left figure) and sequential rank agreement
 (right figure) for comparing two different analysis methods
 (marginal $t$ test and logistic regression)
    applied to the Golub data. The black lines are based on the full
    set of predictors while the red lines have the two highest
    associated predictors removed
    from the data before computing the average overlap and sequential
    rank agreement.}
  \label{fig:case1}
 \end{figure}

 The black lines in Figure~\ref{fig:case1} show the average overlap
 and sequential rank agreement for these data. It is clear from both
 plots that there is perfect agreement towards the top of the ranked
 lists: the average overlap is 1 and the sequential rank agreement is
 zero.  However, if we remove those two items from the data (i.e.,
 gene/predictor 2124 and 896) and redo the analyses then we get
 substantially different curves for the average overlap but not for
 sequential rank agreement (red lines in Figure~\ref{fig:case1}). For
 the sequential rank agreement we get roughly the same estimate of
 agreement from rank 3 as we did from the full dataset. This indicates
 that sequential rank is more robust against small perturbations of
 the data. For the average overlap the curves completely change when
 these two items are removed and which would lead to a different
 conclusion about the agreement among the items from rank 3 downwards.

\section{Application to ovarian cancer data}
We now consider an application of the sequential rank agreement to two
data sets consisting of MALDI-TOF (Matrix-Assisted Laser
Desorption/Ionization Time Of Flight) mass spectra obtained from blood
samples from patients with either benign or malignant ovarian
tumors. The data sets are sub-samples of the Danish MALOVA and DACOVA
study populations.

The MALOVA study is a multidisciplinary Danish study on ovarian cancer
\citep{Hogdall:2004:Cancer:15160342} where all Danish women diagnosed
with an ovarian tumor and referred for surgery from the participating
departments of gynecology were enrolled continuously from December
1994 to May 1999. For the purpose of illustration we use a random
sub-sample of $119$ patients with a total of $58$ patients with
malignant ovarian cancers as cases and $61$ patients with benign
ovarian tumors as controls. The DACOVA study is another
multidisciplinary Danish study on ovarian cancer which included about
$66\%$ of the female population of Denmark
\citep{bertelsen1991protocol}. The study aimed to continuously enroll
all patients that were referred to surgery of an ovarian tumor
clinically suspected to be cancer during the period from 1984 to
1990. Similarly, we use a random sub-sample from the DACOVA study of
$113$ patients with a total of $54$ malignant ovarian cancers and $59$
benign ovarian tumors/gynecologic disorders.

Each spectrum consists of $49642$ samples over a range of mass-to-charge ratios
between $800$ to $20000$ Dalton which we downsample on an equidistant grid of
5000 points by linear interpolation. We then preprocess the downsampled
spectra individually by first removing the slow-varying baseline intensity
with the SNIP algorithm \citep{ryan1988snip} followed by a
normalization with respect to the total ion count. Finally, we standardize
the $5000$ predictors to have column-wise zero mean and unit variance in
each data set.

We use the two data sets to illustrate how the sequential rank
agreement can be applied in two different scenarios. In the first
scenario we assess the agreement of four different statistical
classification methods in how they rank the predictors according to
their importance for distinguishing benign and malignant tumors. In
the second scenario we assess the agreement among rankings of
individual predicted risks of having a malignant tumor. The first
scenario is relevant in the context of biomarker discovery and the
latter is important e.g., when ranking patients according to immediacy of
treatment.

Four classification methods are considered: Random Forest
\citep{breiman2001random} implemented in the R package
\texttt{randomForest} \citep{liaw2002classification}, logistic Lasso
\citep{tibshirani1996regression} and Ridge regression
\citep{segerstedt1992ordinary} both implemented in the R package
\texttt{glmnet} \citep{friedman2010regularization}, and Partial Least
Squares Discriminant Analysis (PLS-DA) \citep{boulesteix2004pls}
implemented in the R package \texttt{caret} \citep{Jed-Wing:2014aa}.

In both scenarios we use the MALOVA data to train the statistical
models, and in both situations the agreements are assessed with
respect to perturbations of the training data in the following
manner. We repeatedly draw a random sub-sample (without replication)
consisting of 90\% of the MALOVA observations and train the four
models on each sub-sample. We use 1000 iterations for the sub-sampling
procedure.

All four methods depend on a tuning parameter. The tuning parameter for
Lasso and Ridge regression is the degree of penalization, and for PLS-DA
it is the number of components (the dimensionality of the subspace).
We estimate these separately for each sub-sample by a 20 times repeated
5-fold cross-validation procedure. For the Random Forest we grow a fixed
number of $5000$ trees and let the tuning parameter be the number of
predictors randomly sampled at each split. We estimate this by a binary
search with respect to minimizing the Out-of-Bag classification error estimate.

The implementation of Lasso and Ridge regression in the
\texttt{glmnet} package offers three different cross-validated
optimization criteria for the penalty parameter: total deviance,
classification accuracy and area under ROC. We apply all three
criteria to our data to investigate their effect on the agreements.
Note also that the Lasso models produce censored lists depending on
the value of the penalty parameter.

\subsection{Agreement of predictor rankings}
For each of the four methods, each of the 1000 models trained on
the 1000 sub-samples of the MALOVA data produces a ranking of the 5000
predictors according to their importance for discriminating between
the tumor types. For the Random Forest classifier the predictors are ranked according
to the Gini index, while for the logistic Lasso and Ridge regression
models we order by absolute magnitude of the estimated
regression coefficients. For the PLS-DA model the importance of the
predictors is based on a weighted sum of the absolute coefficients
where the weights are proportional to the reduction in the sums of
squares across the components.

\begin{figure}[htbp]
\begin{center}
\includegraphics[width=.49\textwidth]{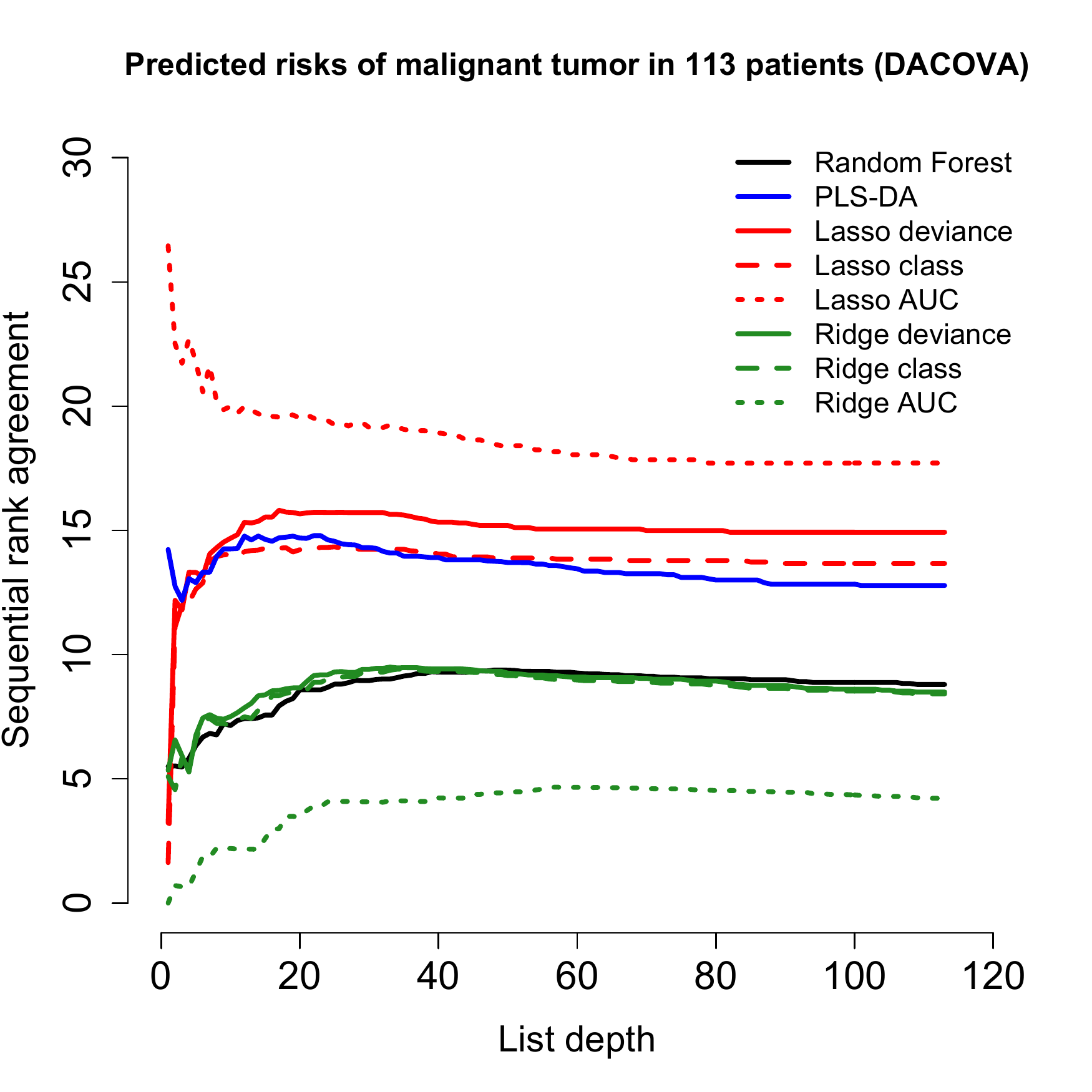}%
\includegraphics[width=.49\textwidth]{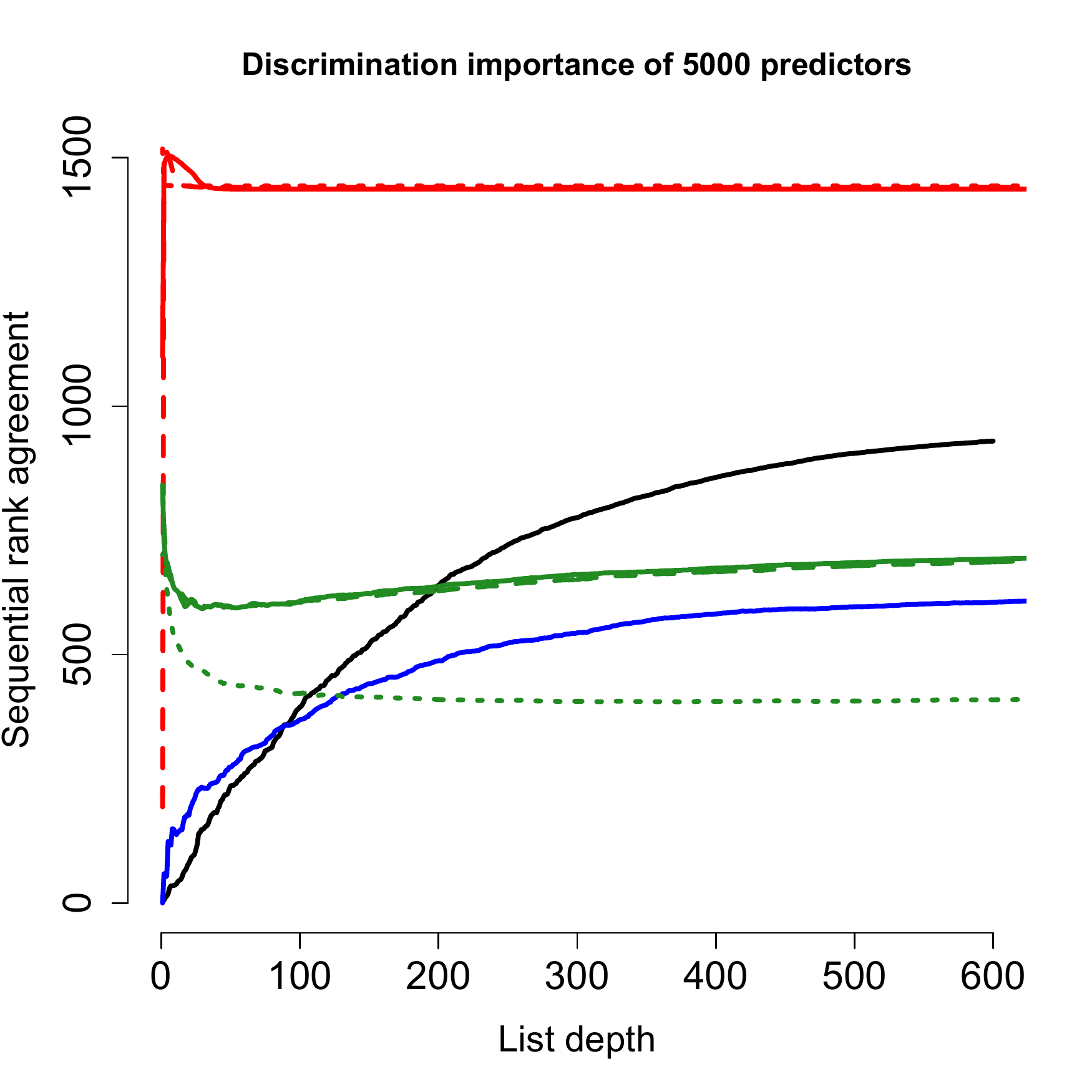}
\end{center}
\caption{Left panel: Sequential rank agreement of 1000 rankings of the
  predicted risks of malignant tumor. For each method the different
  rankings were obtained by first training models in 1000 random
  sub-samples of the MALOVA data and then predicting the risk of
  malignant tumor in the 113 DACOVA patients. Right panel: Sequential
  rank agreement of 1000 rankings of the 5000 predictors. The rankings
  were obtained from the same 1000 trained models.}
 \label{fig:app1}
\end{figure}

The right panel of Figure \ref{fig:app1} shows the sequential rank
agreement of the estimated importance of the 5000 predictors.  For
clarity of presentation we zoom in on the agreement up to list depth
600. At deeper list depths all agreement curves are approximately constant.

For most of the sequential rank agreement curves we see, as expected,
that they start low, indicating good agreement, followed by an
increase until they approximately become constant. This has the
interpretation that the agreement across the different sub-samples is
higher in the top as compared to the tail of the lists for all these
classification methods. The change-points where the curves become
approximately constant are the list depths where the ranks of the
remaining items become close to uniformly random.

A not expected shape of the agreement curves is seen for the Ridge
models for all three tuning criteria. They all show higher
disagreement in the top of the lists followed by a decrease. The
reason behind this behavior is rather subtle. Looking at the
distribution of the absolute value of the regression coefficients we see that a large
proportion of them are numerically very close to zero and have almost
equal absolute value. This is a general feature of the Ridge models in
this data set and seen for all the 1000 trained models.
This implies that when predictors are ranked according to the
magnitude of their coefficients, their actual order becomes more
uncertain and more close to a random permutation. This problem can be
alleviated by truncating all predictors with absolute coefficient values
below a given threshold thereby introducing an artificial censoring of the lists.
For the Ridge models tuned with the deviance criterion, Figure \ref{fig:app2} (left) shows the
the sequential rank agreement where for each of the 1000 trained
models the predictors were artificially censored when their absolute
coefficient value was lower than the $0.1\%$ quantile of the 5000
absolute coefficient values. The curve was calculated
using Algorithm \ref{sra-algorithm} with $B=1000$ and $P=5000$. The corresponding
curve from Figure \ref{fig:app1} (right panel) is shown for comparison.
Even though the number of censored predictors is very small compared to the total
number of predictors, the effect on the sequential rank agreement is
substantial and with the artificial censoring the shape of the curves
is as expected, starting low and then increasing.

Looking at the agreement curves for the Lasso models in Figure \ref{fig:app1}
(right) we clearly see the effect of the sparsity inducing penalization giving
rise to censored lists. These curves were similarly calculated using \ref{sra-algorithm}
and $1000$ random permutations. Under the deviance optimization criterion the
median number of non-zero coefficients was 33 (range 16 to 50) and for the
class accuracy criterion 14 (range 4 to 56). These values correspond to
the list depths where the agreement curves become constant as a result of
the subsequent censoring.

\subsection{Agreement of individual risk predictions}
\label{sec:airp}
To assess the stability of the individual risk predictions we apply
the predictors from the DACOVA data set to each of the models. The
predicted probabilities are then ranked in decreasing order such that the
patients with the highest risk of a malignant tumor appears in the top
of the list. Figure \ref{fig:app1} (left) shows the sequential rank
agreement separately for each method, based on the 1000 risk predictions
obtained from the models trained in the same $1000$ random sub-samples
of the MALOVA data.

Most curves start low and then increase indicating higher agreement
among high risk patients. This is expected if we rank the individuals
according to highest risk of disease. However, it is also expected
that individuals with very low risk also show high agreement. In this
case we order the patients according to (high) risk prediction but we
could essentially also have reversed the order to identify the
patients that have low risk prediction.

An exception is the risk prediction agreement for the Lasso tuned with the AUC criterion which
shows very low agreement among the high values of the predicted
risks. The reason is that optimizing the penalty parameter with
respect to the AUC criterion tends to favor a very high penalty value
causing only a single predictor to be selected in each of the 1000
iterations. This results in a lack of generalizability to the DACOVA
data which gives rise to the higher disagreement in the predicted risks.
In the extreme case where the penalty becomes so high
that none of the predictors are selected by the Lasso, the sequential
rank agreement for the predicted probabilities becomes undefined since
all the ranks will be ties.

\begin{figure}[htbp]
\begin{center}
\includegraphics[width=.49\textwidth]{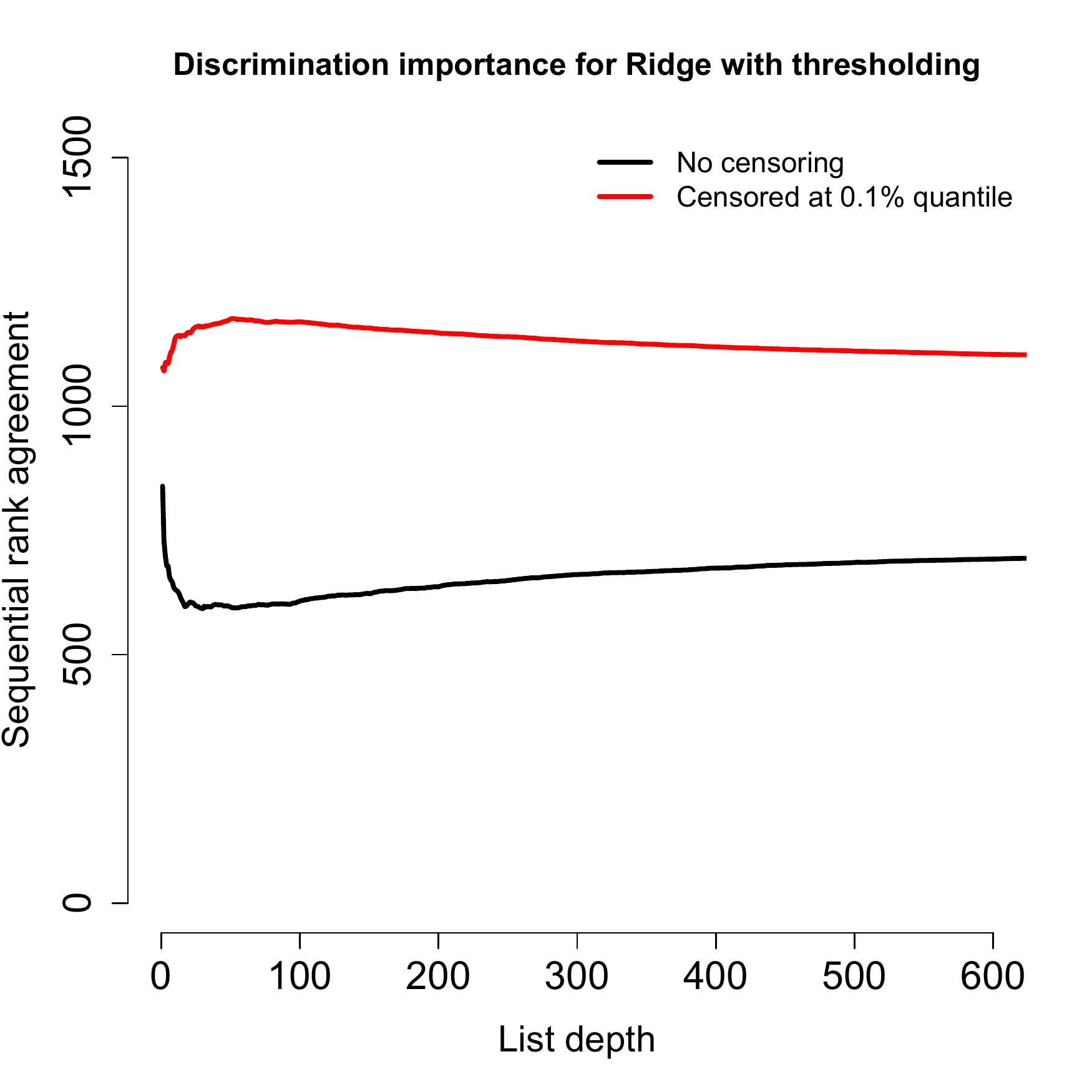}%
\includegraphics[width=.49\textwidth]{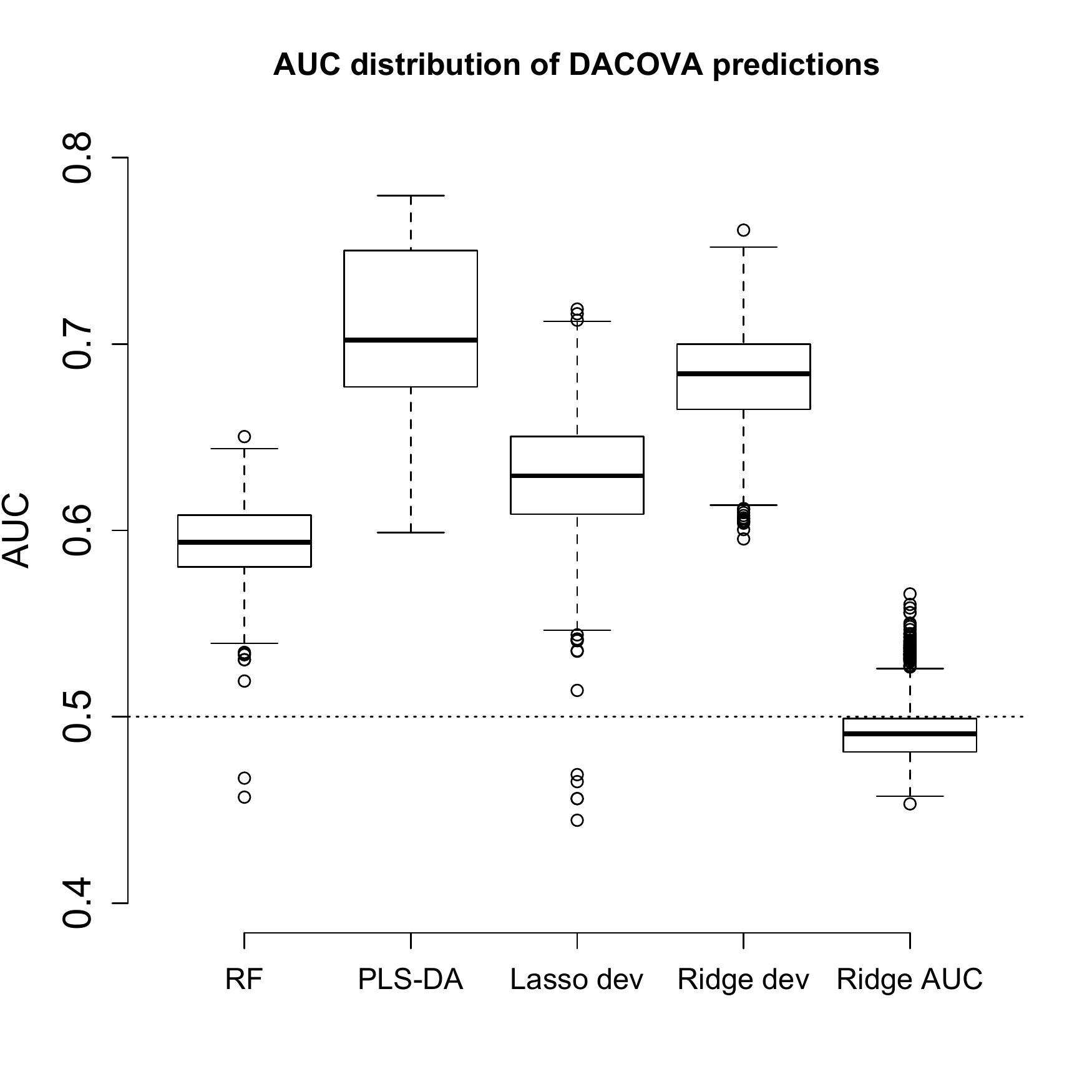}
\end{center}
\caption{Left panel: Sequential rank agreement for Ridge regression
  obtained by artificially censoring predictors when their absolute
  coefficient values are lower than the 0.1\% quantile. Right panel:
  Box plots of AUC values across the $1000$ sub-samples with respect
  to the known class labels of the DACOVA data.}
 \label{fig:app2}
\end{figure}

Comparing the left and right panels of Figure \ref{fig:app1} it can
further be seen that some of the methods show better agreement with
respect to the predicted probabilities than for ranking the importance
of the predictors and vice versa. Ridge regression shows higher agreement across training
sets for the risk predictions than PLS-DA, and PLS-DA shows higher
agreement for predictor importance than Ridge regression.

Lasso shows similar agreement for ranking the risk predictions than
PLS-DA (unless with the AUC criterion), and poorer agreement for ranking
predictors. This reason for the latter is the high auto-correlation
between the intensities in the mass spectra which leads to
collinearity issues in the regression models. It is well-known that
variable selection with the Lasso penalty does not preform very well
when the predictors are highly correlated. The collinearity does,
however, not affect the agreement of the risk predictions, as seen on
the left panel of Figure \ref{fig:app1}, since it is not so important
which specific variable that gets selected from a group of highly correlated
predictors when the purpose is risk predictions.

It appears that Ridge regression tuned with the AUC criterion achieves
the best performance with respect to the stability of ranking the
individual predicted risk probabilities. It must, however, be stressed
that the sequential rank agreement in this application is only
concerned with the agreement of the risk predictions across
sub-samples and not with the actual accuracy of the risk
predictions. Thus, we also computed the AUC values for the different models
based on the DACOVA data. The distributions across the $1000$
sub-samples for a selection of the models is shown in the right panel
of Figure \ref{fig:app2}.  Here we see that PLS-DA attains the highest
AUC values with a median value of $0.70$ while the Ridge model with
the AUC criterion attains a median AUC of $0.49$. This implies that
while Ridge regression optimized with respect to the AUC criterion
achieves the best sequential rank agreement, it performs similar to a
random coin toss with respect to classifying the DACOVA patients.
In practice both concerns are of importance.

\section{Discussion}
In this article we address the problem of comparing ranked lists of
the same items. Our proposed method can handle both the situation
where the underlying data to generate the ranked lists are available
and the situation where the only available data is the actual ranked
lists. In addition, censored ranked lists where only the ranks of the
top $k$ ranked items are known can be accommodated as well. The
proposed agreement measure can be interpreted as the average distance
between an item's rank and the average rank assigned to that item
across lists. Thus the measure determines how well the lists agree on
the ranks of a specific item.

The sequential rank agreement is extremely versatile. We have shown
that it can be used not only to compare ranked lists of items produced
from different samples/populations but that it also can be used to
study the ranks obtained from different analysis methods on the same
data as well as to evaluate the stability of the ranks from a single
method by bootstrapping (or sub-sampling) the data repeatedly and
comparing the ranks obtained from training the models in the
bootstrapped data.

The sequential rank agreement can be used to determine the depth at
which the rank agreement becomes too large to be desirable based on
prior requirements or acceptable differences, or it can be used to
visually determine when the change in agreement becomes too large.  In
that regard the investigator can have prior limits on the level of
agreement that is acceptable.

While the sequential rank agreement is primarily an exploratory tool
we have suggested two null hypotheses that can be used to evaluate the
sequential rank agreement obtained. Note that none of the two null
hypotheses are concerned with the actual ``true ranking'' but are
purely concerned with consistency/stability of the rankings among the
lists. As such we cannot determine if the rankings are good but only
whether they agree. The sequential rank agreement curve can be
compared visually to the curves obtained under either of the null
distributions and simple point-wise $p$-values can be obtained for
each depth by counting the number of sequential rank agreements under
the null hypothesis that is less than or equal to the observed rank
agreement. Two simple extensions can be pursued to make more formal
uniform tests: one would be to use a Kolmogorov-Smirnov-like test and
use the largest difference between the variance-weighted sequential
rank agreement curve and the mean null rank agreement curve as a test
statistic. Alternatively, a change-point analysis could be made on the
sequential rank agreement in order to determine the depths at which
there are ``jumps'' in the rank agreement. These jumps would
correspond to depths for which the agreement among the lists was
substantially worse and could serve as indicators for when the lists
no longer agree sufficiently satisfactory.

Finally, we have --- whenever possible --- used all available ranks
from the lists.  We could choose to censor the rankings of all the
items that were deemed not to be significant if information on both
rankings and the corresponding evidence for significance was
available. That would ensure that there would be put less emphasis on
the agreement of the non-significant items and it would be easier to
identify a change in agreement among the items that were deemed to
be relevant. In our application section we have successfully introduced
such an artificial censoring for the predictor rankings obtained with
ridge regression.

We note that the sequential rank agreement is still marred by problems
that generally apply to ranking of items and/or
individuals. Collinearity in particular can be a huge problem when
bootstrapping data or when comparing different analysis methods. For
example, marginal analyses where each item is analyzed separately will
assign similar ranks to two highly correlated predictors while methods
that provide a sparse solution such as the Lasso will just rank one of
the two predictors high while the other might have a very low rank.
Thus in such a scenario we would expect low agreement of the rankings
from Lasso and marginal analyses simply because of the way correlated
predictors are handled. This is not a shortcoming of the sequential
rank agreement but is a problem general to all ranked lists.

Another caveat with the way the sequential rank agreement is defined
is the use of the standard deviation to measure agreement. The
standard deviation is an integral part of the limits-of-agreement as
discussed by \citet{alt:bland:1983}. However, the standard deviation
can also be unstable when the number of observations is low and
alternative measures such as the median absolute deviance may prove more
stable in some situations. However, the current definition using the
standard deviation is analogous to the approach used for agreement in
method comparison studies so we have used that.

In conclusion we have introduced a method for evaluation of ranked
(censored) lists that can be easily interpreted and that can be
applied to a large number of situations.  The method presented here
can be adapted further by using it to compare and classify statistical
analysis methods that agree on the rankings they provide or by using
the rank agreement to optimize a hyper-parameter in, say, elastic net
regularized regression where the rank agreement is used to determine
the mixing proportion between the $L_1$ and the $L_2$ penalty.
We will be investigating these extensions further in the future.

\bibliographystyle{rss}
\bibliography{paperref}

\end{document}